\documentclass[conference]{IEEEtran}
\IEEEoverridecommandlockouts
\usepackage{mathrsfs}
\usepackage{graphicx}
\usepackage{float}
\usepackage{hyperref}
\usepackage{cite}

\usepackage{enumitem}
\usepackage{color}
\usepackage{psfrag}
\usepackage{subfigure}
\usepackage{amssymb}
\usepackage{amsthm}
\usepackage{setspace}
\usepackage{epsfig}
\usepackage{pifont}
\usepackage{amsmath}
\usepackage{array}
\usepackage{multicol}
\usepackage{multirow}
\usepackage{diagbox}
\usepackage{pifont}
\usepackage{indentfirst}
\usepackage{amsfonts}
\usepackage{algorithm}
\usepackage{algorithmicx}
\usepackage{algpseudocode}
\usepackage{fancyhdr}
\usepackage{amscd}
\usepackage{bm}
\usepackage{fancyhdr}
\usepackage{enumerate}
\usepackage{color}
\usepackage{threeparttable}
\renewcommand{\baselinestretch}{0.97}

\newtheorem{proposition}{Proposition}

\IEEEoverridecommandlockouts

\makeatletter
\renewcommand{\citepunct}{,\penalty\@m\hskip.13emplus.1emminus.1em}
\renewcommand{\citedash}{\hbox{--}\penalty\@m}

\begin{document}
\title{Precoder Learning for Weighted Sum Rate~Maximization}
		
\author{\IEEEauthorblockN{Mingyu Deng and Shengqian Han}
	\IEEEauthorblockA{School of Electronics and Information Engineering, Beihang University, Beijing 100191, China\\
		Email: \{mingyudeng, sqhan\}@buaa.edu.cn}
}
\maketitle

\begin{abstract}
Weighted sum rate maximization (WSRM) for precoder optimization effectively balances performance and fairness among users. Recent studies have demonstrated the potential of deep learning in precoder optimization for sum rate maximization. However, the WSRM problem necessitates a redesign of neural network architectures to incorporate user weights into the input. In this paper, we propose a novel deep neural network (DNN) to learn the precoder for WSRM. Compared to existing DNNs, the proposed DNN leverage the joint unitary and permutation equivariant property inherent in the optimal precoding policy, effectively enhancing learning performance while reducing training complexity. Simulation results demonstrate that the proposed method significantly outperforms baseline learning methods in terms of both learning and generalization performance while maintaining low training and inference complexity.
\end{abstract}


\section{Introduction}
In multiuser multi-antenna systems, optimizing precoders for weighted sum rate maximization (WSRM) can achieve proportional fairness among user equipments (UEs)~\cite{proportional_fairness2}. Various numerical algorithms, such as the weighted minimum mean square error (WMMSE) algorithm~\cite{WMMSE}, have been developed to address WSRM problems. While these methods can achieve high performance, they often incur significant complexity due to iterative~computations.

To achieve low inference time while maintaining satisfactory learning performance, deep learning methods have been increasingly utilized for learning wireless resource management strategies~\cite{guo_multi-cell-multi-user,shiqingjiang_DNN}. In particular, the effectiveness of graph neural networks (GNNs) for precoder learning has gained considerable attention~\cite{Eisen2020,He2021,shen_yifri_framework}. Compared to traditional deep neural networks (DNNs), GNNs offer advantages primarily through the incorporation of appropriate inductive biases into their architecture~\cite{inductive_biase_gnn,edgegnn}, enabling them to align with the permutation equivariant (PE) properties of the policies to be learned.

When designing the DNN structure, the incorporation of the same mathematical property as the wireless policy can reduce the hypothesis space, and thus improve the learning efficiency~\cite{edgegnn}. Take the PE property for instance, in \cite{edgegnn}, a heterogeneous Edge-GNN was designed to satisfy the PE property of the precoding policy regarding the indices of UEs and antennas, i.e., the changes in the order of these indices in the input will result in corresponding changes in the output. Thus compared to traditional fully connected and conventional neural networks, Edge-GNN gets better learning performance with given training set and lower training complexity to achieve an expected performance.

When designing a network that satisfies mathematical properties, the focus is primarily on designing the weight matrix of weight function in each hidden layer. The more and stronger mathematical properties weight matrices satisfy, the more the hypothesis space can be reduced. For example, existing research indicates that introducing PE property leads to a parameter sharing mechanism in the linear weight matrices~\cite{jointPE,liu2023multidimensional}, which reduces the number of free parameters. Furthermore, as the strength of PE property increases, the number of free parameters tends to decrease, thereby improving learning efficiency. For example, the PE property regarding the indices of both antennas and UEs results in fewer free parameters than considering just one type of index~\cite{guo_multi-cell-multi-user,liu2023multidimensional}. In addition to PE property, some studies have designed nonlinear weight matrices to introduce other mathematical properties. The nonlinear weight matrices designed in \cite{model_gnn} and \cite{gradient_gnn} introduce the mathematical properties of the Taylor expansion of the matrix pseudo-inverse and the gradient of the sum rate, respectively, both of which enhance the training efficiency and learning performance.

While existing studies have extensively explored the PE property for sum rate maximization (SRM) problems, they often overlook a stronger equivariant property, i.e., the unitary equivariant property. This property acts on the antenna dimension, where applying a unitary transformation to the multi-antenna channel vector of each UE results in a corresponding transformation of the precoder. Notably, since permutation matrices are a specific type of unitary matrix, the unitary equivariant property is stronger than the PE property, offering potential benefits for improving learning performance and reducing training complexity. However, the application of the unitary property for learning the precoder in WSRM problems has yet to be addressed.

In this paper, we design a DNN to learn the precoding policy for the WSRM problem in a multiuser multi-antenna system. Existing DNNs designed for SRM problems are not suitable for WSRM, as the latter requires inputting UE weights into the network alongside channel information.
We first derive the structure of the linear weight matrix in each hidden layer that satisfies both the unitary equivariant property regarding channel vectors and the PE property regarding the indices of UEs. We then demonstrate that this weight matrix cannot incorporate UE weights into the DNN, and the loss of this mathematical property instead leads to the inability to learn the optimal precoding policy. To address this, we propose a nonlinear weight matrix that effectively leverages the joint equivariant property. Simulation results demonstrate the advantages of our proposed method in terms of learning and generalization performance while reducing sample and time complexity.

\section{System Model and Equivariant Property}

Consider a downlink transmission where a base station (BS) equipped with $N$ antenna serves $K$ single-antenna UEs. Let ${{\mathbf{h}}_{k}}\in {{\mathbb{C}}^{N\times 1}}$ and ${{\mathbf{v}}_{k}}\in {{\mathbb{C}}^{N\times 1}}$ denote the channel vector and precoder of UE$_k$ during a time slot, and $\mathbf{H}=\left[ {{\mathbf{h}}_{1}},\ldots ,{{\mathbf{h}}_{K}} \right]\in {{\mathbb{C}}^{N\times K}}$ and $\mathbf{V}=\left[ {{\mathbf{v}}_{1}},\ldots ,{{\mathbf{v}}_{K}} \right]\in {{\mathbb{C}}^{N\times K}}$ denote the overall channel and precoder matrices for all UEs, respectively.

To ensure proportional fairness among the UEs, the following WSRM problem is solved in every time slot:

\begin{subequations}\label{WSMR_problem2}
    \begin{align}
        & \max \text{  }\sum\limits_{k=1}^{K}{{{\alpha }_{k}}\log \left( 1+\frac{{{\left| \mathbf{h}_{k}^{\text{H}}{{\mathbf{v}}_{k}} \right|}^{2}}}{\sum\limits_{i\ne k}{{{\left| \mathbf{h}_{k}^{\text{H}}{{\mathbf{v}}_{i}} \right|}^{2}}+{{\delta }^{2}}}} \right)} \label{E:obj}  \\
        &\ \text{ s.t.}\text{    }\sum\limits_{k=1}^{K}{{{\left| {{{\bf v}}_{k}} \right|}^{2}}}\le {{P}_{m}},
    \end{align}
\end{subequations}
where $\alpha_k$ is the weight assigned to UE$_k$ in the current time slot, which can be selected as the inverse of the average rate of UE$_k$ over the past slots, ${{\delta }^{2}}$ is the noise power, and $P_m$ is the total transmit power of the BS.

The optimal precoding policy is a mapping from channels $\mathbf{H}$ and UE weights $\mathbf{\Lambda}={\left[ {{\alpha }_{1}},\ldots ,{{\alpha }_{K}} \right]^T}\in {{\mathbb{R}}^{K\times 1}}$ to the optimal precoder $\mathbf{V}$, denoted by $\mathbf{V}=\mathcal{F}\left( \mathbf{H}, \mathbf{\Lambda } \right)$. The equivariant property of the precoding policy can be analyzed as follows.

The rows and columns of $\mathbf{H}$ correspond to antenna and UE dimensions, respectively. If the antenna dimension of $\mathbf{H}$ is transformed by an arbitrary unitary matrix $\mathbf{U}\in {{\mathbb{C}}^{N\times N}}$, i.e., left-multiplying $\mathbf{H}$ by $\mathbf{U}$, and simultaneously, the indices of UEs in both $\mathbf{H}$ and $\mathbf{\Lambda}$ are permutated, i.e., right-multiplying $\mathbf{H}$ and $\mathbf{\Lambda}^T$ by an arbitrary permutation matrix $\mathbf{\Pi}^T\in {{\mathbb{R}}^{K\times K}}$, it can be found that transforming the optimal precoder $\mathbf{V}$ in the same way does not change the value of the objective function in \eqref{E:obj}. Such a joint equivariant property can be expressed as
\begin{equation}\label{optimal_function}
    \mathbf{UV}{{\mathbf{\Pi }}^{\text{T}}}=\mathcal{F}\left( \mathbf{UH}{{\mathbf{\Pi }}^{\text{T}}},\mathbf{\Lambda}^T\mathbf{\Pi}^T \right).
\end{equation}

Our objective is to design a DNN that satisfies the same equivariant property as the precoding policy $\mathcal{F}(\cdot)$.

\section{Design of Equivariant DNN}

The design of an equivariant DNN can be achieved by ensuring that every hidden layer satisfies the equivariant property~\cite{jointPE}.
This involves two issues: identifying the input and output for each hidden layer, and ensuring that the hidden input and output satisfy the equivariant property. We next address these two issues, respectively.

\subsection{Hidden Input and Output}

Regarding the precoder learning for SRM, existing works have investigated the selection of hidden input and output based on the WMMSE algorithm. In \cite{RGNN}, the iterative equation of the WMMSE algorithm is interpreted as the update equation of a GNN. From this perspective, for the WSRM problem, it can be found that the precoder is updated based on the precoder from the previous iteration, channel information, and the UE weights. Therefore, for the $l$-th hidden layer, we define the output to have the same dimension as the precoder $\mathbf{V}$, denoted as ${{\mathbf{X}}^{\left( l \right)}}=\left[ \mathbf{x}_{1}^{\left( l \right)},\ldots ,\mathbf{x}_{K}^{\left( l \right)} \right]\in {{\mathbb{C}}^{N\times K}}$. The input includes ${{\mathbf{X}}^{\left( l-1 \right)}}$, $\mathbf{H}$, and $\mathbf{\Lambda}$.
The output ${{\mathbf{X}}^{\left( l \right)}}$ is also referred to as the hidden features. For simplicity, we consider a single hidden feature channel for each layer. However, the proposed method can be readily extended to accommodate multiple hidden feature channels, similar to traditional DNNs.

To facilitate the expression of the input-output relationship for each hidden layer, we represent the input and output in vectorized format. Specifically, we define ${{\overrightarrow{\mathbf{X}}}^{\left( l \right)}}=\text{vec}\left( {{\mathbf{X}}^{\left( l \right)}} \right)\in {{\mathbb{C}}^{NK\times 1}}$. For the input layer, we set ${{\overrightarrow{\mathbf{X}}}^{\left( 0 \right)}}=\overrightarrow{\mathbf{H}}$. The output of the final layer corresponds to the learned precoder, denoted as $\overrightarrow{\mathbf{V}}={{\overrightarrow{\mathbf{X}}}^{\left( L \right)}}$, where $L$ is the total number of layers in the~DNN.

The relationship between the input and output of the $l$th layer can be expressed as
\begin{equation}\label{approximate_optimal_function}
    {{\overrightarrow{\mathbf{X}}}^{\left( l \right)}}=\sigma^{\left( l \right)} \left( \psi^{\left( l \right)} \left( {{\overrightarrow{\mathbf{X}}}^{\left( l-1 \right)}},\overrightarrow{\mathbf{H}},\mathbf{\Lambda } \right) \right),
\end{equation}
where $\psi^{\left( l \right)} \left( \cdot  \right)$ is a weight function, and $\sigma^{\left( l \right)} \left( \cdot  \right)$ is an activation function. To approximate the optimal precoding policy $\mathcal{F}\left(\cdot\right)$, both $\psi^{\left( l \right)} \left( \cdot  \right)$ and $\sigma^{\left( l \right)} \left( \cdot  \right)$ should satisfy the joint equivariant property.
The activation function $\sigma^{\left( l \right)} \left( \cdot  \right)$ can be appropriately selected, as will be discussed later.

In the sequel, we temporally disregard $\sigma^{\left( l \right)} \left( \cdot  \right)$ and focus on the design of the weight function $\psi^{\left( l \right)} \left( \cdot  \right)$.

\subsection{Linear weight Function}
A widely used weight function $\psi^{\left( l \right)} \left( \cdot  \right)$ for DNNs is the linear weight function. By temporally omitting  $\sigma^{\left( l \right)} \left( \cdot  \right)$, we can rewrite \eqref{approximate_optimal_function} with a linear $\psi^{\left( l \right)} \left( \cdot  \right)$ as
\begin{equation}\label{static_linear_weight_matrix1}
    {{\overrightarrow{\mathbf{X}}}^{\left( l \right)}}=\mathbf{W}^{\left( l \right)}{{\left[ {{\left( {{\overrightarrow{\mathbf{X}}}^{\left( l-1 \right)}} \right)}^{\text{T}}},{{\overrightarrow{\mathbf{H}}}^{\text{T}}},{{\mathbf{\Lambda }}^{\text{T}}} \right]}^{\text{T}}},
\end{equation}
where $\mathbf{W}^{\left( l \right)}\in {{\mathbb{C}}^{NK\times \left( 2NK+K \right)}}$ is the weight matrix. We can partition $\mathbf{W}^{\left( l \right)}$ into three sub-matrices as $\mathbf{W}^{\left( l \right)}=\left[ \mathbf{W}_{1}^{\left( l \right)},\mathbf{W}_{2}^{\left( l \right)},\mathbf{W}_{3}^{\left( l \right)} \right]$. Then, \eqref{static_linear_weight_matrix1} can be rewritten as
\begin{equation}\label{static_linear_weight_matrix2}
    {{\overrightarrow{\mathbf{X}}}^{\left( l \right)}}=\mathbf{W}_{1}^{\left( l \right)}{{\overrightarrow{\mathbf{X}}}^{\left( l-1 \right)}}+\mathbf{W}_{2}^{\left( l \right)}\overrightarrow{\mathbf{H}}+\mathbf{W}_{3}^{\left( l \right)}\mathbf{\Lambda },
\end{equation}
where  $\mathbf{W}_{1}^{\left( l \right)}\in {{\mathbb{C}}^{NK\times NK}}$, $\mathbf{W}_{2}^{\left( l \right)}\in {{\mathbb{C}}^{NK\times NK}}$ and $\mathbf{W}_{3}^{\left( l \right)}\in {{\mathbb{C}}^{NK\times K}}$.

We first examine the conditions that the three sub-matrices must satisfy to fulfill the unitary equivariant property. This property requires that when the input ${{\mathbf{X}}^{\left( l-1 \right)}}$ and $\mathbf{H}$ are transformed into ${\mathbf{U}{\mathbf{X}}^{\left( l-1 \right)}}$ and $\mathbf{U}\mathbf{H}$, the output should become ${\mathbf{U}{\mathbf{X}}^{\left( l \right)}}$. Based on the property of Kronecker product, we can derive the following condition
\begin{equation}\label{static_linear_weight_matrix3}
    \begin{aligned}
         &\left({{\mathbf{I}}_{K}}\otimes \mathbf{U}\right){{\overrightarrow{\mathbf{X}}}^{\left( l \right)}} \\
         &=\mathbf{W}_{1}^{\left( l \right)}{\left({{\mathbf{I}}_{K}}\!\otimes \! \mathbf{U}\right)}{{\overrightarrow{\mathbf{X}}}^{\left( l-1 \right)}}\!+\!\mathbf{W}_{2}^{\left( l \right)}\left({{\mathbf{I}}_{K}}\!\otimes \! \mathbf{U}\right)\overrightarrow{\mathbf{H}}\!+\!\mathbf{W}_{3}^{\left( l \right)}{{\mathbf{\Lambda }}^{\text{T}}},
    \end{aligned}
\end{equation}

Substitute \eqref{static_linear_weight_matrix2} into \eqref{static_linear_weight_matrix3}, we yield
\begin{equation}\label{static_linear_weight_matrix4}
    \begin{aligned}
        &\left({{\mathbf{I}}_{K}}\otimes \mathbf{U}\right)\left( \mathbf{W}_{1}^{\left( l \right)}{{\overrightarrow{\mathbf{X}}}^{\left( l-1 \right)}}+\mathbf{W}_{2}^{\left( l \right)}\overrightarrow{\mathbf{H}}+\mathbf{W}_{3}^{\left( l \right)}\mathbf{\Lambda } \right) \\
        &=\mathbf{W}_{1}^{\left( l \right)}{\left({{\mathbf{I}}_{K}}\!\otimes \! \mathbf{U}\right)}{{\overrightarrow{\mathbf{X}}}^{\left( l-1 \right)}}\!+\!\mathbf{W}_{2}^{\left( l \right)}\left({{\mathbf{I}}_{K}}\!\otimes \! \mathbf{U}\right)\overrightarrow{\mathbf{H}}\!+\!\mathbf{W}_{3}^{\left( l \right)}{{\mathbf{\Lambda }}^{\text{T}}}.
    \end{aligned}
\end{equation}

The condition in \eqref{static_linear_weight_matrix4} for the three sub-matrices must be satisfied for arbitrary input  ${{\mathbf{X}}^{\left( l-1 \right)}}$, $\mathbf{H}$, and $\mathbf{\Lambda}$. To analyze this, we consider two special cases for the input:
\begin{itemize}
  \item\emph{Case 1: $\overrightarrow{\mathbf{H}}=\mathbf{0}$ and $\mathbf{\Lambda}=\mathbf{0}$.}

  In this case, we can derive a condition for the sub-matrix $\mathbf{W}_{1}^{\left( l \right)}$ from \eqref{static_linear_weight_matrix4} as
\begin{equation}\label{static_linear_weight_matrix433}
    \begin{aligned}
        &\left({{\mathbf{I}}_{K}}\otimes \mathbf{U}\right)\mathbf{W}_{1}^{\left( l \right)}=\mathbf{W}_{1}^{\left( l \right)}{\left({{\mathbf{I}}_{K}}\!\otimes \! \mathbf{U}\right)},
    \end{aligned}
\end{equation}
which must hold for any arbitrary unitary matrix $\mathbf{U}$. It can be shown that the sub-matrix $\mathbf{W}_{1}^{\left( l \right)}$ should take the following structure (the proof is omitted due to space constraints)
\begin{equation}\label{vector_unitary_equivariant2}
    \mathbf{W}_{1}^{\left( l \right)}=\left(\mathbf{G}_{1}^{\left( l \right)}\right)^T\otimes {{\mathbf{I}}_{N}},
\end{equation}
where $\mathbf{G}_{1}^{\left( l \right)}$ is an arbitrary matrix.

Similarly, we can show that the sub-matrix $\mathbf{W}_{2}^{\left( l \right)}$ has the same structure as given by \eqref{vector_unitary_equivariant2}, which can be expressed~as
\begin{equation}\label{vector_unitary_equivariant23}
    \mathbf{W}_{2}^{\left( l \right)}=\left(\mathbf{G}_{2}^{\left( l \right)}\right)^T\otimes {{\mathbf{I}}_{N}},
\end{equation}
where $\mathbf{G}_{2}^{\left( l \right)}$ is another arbitrary matrix.

\end{itemize}

\begin{itemize}
  \item\emph{Case 2: ${{\overrightarrow{\mathbf{X}}}^{\left( l-1 \right)}}=\mathbf{0}$ and $\overrightarrow{\mathbf{H}}=\mathbf{0}$.}

      In this case, we can derive a condition for the sub-matrix $\mathbf{W}_{3}^{\left( l \right)}$ from \eqref{static_linear_weight_matrix4} as
\begin{equation}\label{static_linear_weight_matrix5}
    \left( \left({{\mathbf{I}}_{K}}\otimes \mathbf{U}\right)\mathbf{W}_{3}^{\left( l \right)}-\mathbf{W}_{3}^{\left( l \right)} \right)\mathbf{\Lambda }=\mathbf{0}.
\end{equation}
Since \eqref{static_linear_weight_matrix5} must hold for any arbitrary $\mathbf{\Lambda }$, it follows that $\mathbf{W}_{3}^{\left( l \right)}$ must satisfy
\begin{equation}\label{static_linear_weight_matrix6}
    \left({{\mathbf{I}}_{K}}\otimes \mathbf{U}\right)\mathbf{W}_{3}^{\left( l \right)}=\mathbf{W}_{3}^{\left( l \right)}.
\end{equation}

To determine the structure of $\mathbf{W}_{3}^{\left( l \right)}$ that meets the condition in \eqref{static_linear_weight_matrix6}, we can divide $\mathbf{W}_{3}^{\left( l \right)}$ into $K$ sub-matrices~as
\begin{equation} \label{E:W3}
    \mathbf{W}_{3}^{\left( l \right)}={{\left[ {{\left( \mathbf{W}_{3,1}^{\left( l \right)} \right)}^{\text{T}}},\cdots ,{{\left( \mathbf{W}_{3,K}^{\left( l \right)} \right)}^{\text{T}}} \right]}^{\text{T}}},
\end{equation}
where $\mathbf{W}_{3,k}^{\left( l \right)}\in {{\mathbb{C}}^{N\times K}},k=1,\cdots ,K$.
By substituting \eqref{E:W3} into \eqref{static_linear_weight_matrix6}, we find that $\mathbf{W}_{3,k}^{\left( l \right)}=\mathbf{UW}_{3,k}^{\left( l \right)}$, $\forall k$.
Note that this condition must hold for any arbitrary unitary matrix $\mathbf{U}$. Considering a special unitary matrix $\mathbf{U}=-\mathbf{I}_N$, we obtain $\mathbf{W}_{3,k}^{\left( l \right)}=-\mathbf{W}_{3,k}^{\left( l \right)}$, which indicates that $\mathbf{W}_{3,k}^{\left( l \right)}$ must be zero for all $k$.
\end{itemize}

By substituting \eqref{vector_unitary_equivariant2}, \eqref{vector_unitary_equivariant23}, and \eqref{E:W3} into \eqref{static_linear_weight_matrix2}, and converting the vectorized result into matrix format, we obtain
\begin{equation}\label{linearResult}
    {{{\mathbf{X}}}^{\left( l \right)}}={{{\mathbf{X}}}^{\left( l-1 \right)}}\mathbf{G}_{1}^{\left( l \right)} + {\mathbf{H}}\mathbf{G}_{2}^{\left( l \right)}+\mathbf{0} \cdot \mathbf{\Lambda}.
\end{equation}
It indicates that, when designing the linear weight matrices, the attempt to introduce the stronger unitary equivariant property instead results in the loss of another important mathematical property: the UE weights $\mathbf{\Lambda }$ cannot be integrated into the weight function of each layer. Consequently, this prevents the learning of the optimal precoder for the WSRM problem.

\subsection{Nonlinear Weight Function}
The analysis in the previous subsection indicates that by employing linear weight function the DNN is unable to learn the optimal precoder while satisfy the joint equivariant property. Therefore, in this subsection, we strive to design a nonlinear weight function to address the problem.

\subsubsection{\underline{Form of nonlinear weight function}}

Inspired by the structure presented in \eqref{linearResult}, and with the goal of integrating $\mathbf{\Lambda}$ into the update equation of each layer, we propose the following weight function
\begin{equation}\label{adaptive_matrix_form0}
    {{\mathbf{X}}^{\left( l \right)}}={{\mathbf{X}}^{\left( l-1 \right)}}\mathcal{G}_{1}^{\left( l \right)}(\mathbf{\Lambda}) + {\mathbf{H}}\mathcal{G}_{2}^{\left( l \right)}(\mathbf{\Lambda}),
\end{equation}
where, instead of using two constant weight matrices $\mathbf{G}_{1}^{\left( l \right)}$ and $\mathbf{G}_{2}^{\left( l \right)}$ as in \eqref{linearResult}, we replace them with two functions $\mathcal{G}_{1}^{\left( l \right)}(\mathbf{\Lambda})$ and $\mathcal{G}_{2}^{\left( l \right)}(\mathbf{\Lambda})$, both of which depend on $\mathbf{\Lambda}$.

The equation \eqref{adaptive_matrix_form0} can be rewritten as
\begin{equation}\label{adaptive_matrix_form1}
    {{\mathbf{X}}^{\left( l \right)}}={{\mathbf{X}}^{\left( l-1 \right)}}{{ \overline{\mathcal{G}}^{\left( l \right)}\left( {{\mathbf{X}}^{\left( l-1 \right)}}, \mathbf{H}, \mathbf{\Lambda} \right)}},
\end{equation}
where nonlinear weight matrix ${{ \overline{\mathcal{G}}^{\left( l \right)}\left( {{\mathbf{X}}^{\left( l-1 \right)}}, \mathbf{H}, \mathbf{\Lambda} \right)}}$ is a function of ${{\mathbf{X}}^{\left( l-1 \right)}}$, $\mathbf{H}$, and $\mathbf{\Lambda}$. The equivalence between \eqref{adaptive_matrix_form0} and \eqref{adaptive_matrix_form1} can be established by setting ${{ \overline{\mathcal{G}}^{\left( l \right)}\left( {{\mathbf{X}}^{\left( l-1 \right)}}, \mathbf{H}, \mathbf{\Lambda} \right)}} = \mathcal{G}_{1}^{\left( l \right)}(\mathbf{\Lambda}) + \left({{\mathbf{X}}^{\left( l-1 \right)}}\right)^{-1}{\mathbf{H}}\mathcal{G}_{2}^{\left( l \right)}(\mathbf{\Lambda})$.

It is evident that the relationship between ${\mathbf{X}}^{\left( l \right)}$ and ${\mathbf{X}}^{\left( l-1 \right)}$ in \eqref{adaptive_matrix_form1} is nonlinear. The following proposition describes the form of ${{ \overline{\mathcal{G}}^{\left( l \right)}\left( {{\mathbf{X}}^{\left( l-1 \right)}}, \mathbf{H}, \mathbf{\Lambda} \right)}}$.
\begin{proposition}\label{input_of_function_of_weight_matrix}
The function ${{ \overline{\mathcal{G}}^{\left( l \right)}\left( {{\mathbf{X}}^{\left( l-1 \right)}}, \mathbf{H}, \mathbf{\Lambda} \right)}}$ can be expressed in terms of the elements $\mathbf{h}_{i}^{\text{H}}\mathbf{x}_{j}^{\left( l-1 \right)}$ and ${{\alpha }_{i}}\mathbf{h}_{i}^{\text{H}}\mathbf{x}_{j}^{\left( l-1 \right)}$ for $i,j=1,\dots,K$, i.e.,
\begin{align} \label{E:Gbar}
  {{ \overline{\mathcal{G}}^{\left( l \right)}\left( {{\mathbf{X}}^{\left( l-1 \right)}}, \mathbf{H}, \mathbf{\Lambda} \right)}} = {{ \mathsf{\mathcal{G}}^{\left( l \right)}\left( \mathbf{E}_1, \mathbf{E}_2 \right)}},
\end{align}
where $\mathsf{\mathcal{G}}^{\left( l \right)}(\cdot)$ is a function of $\mathbf{E}_1, \mathbf{E}_2\in\mathbb{C}^{K\times K}$, the $(i,j)$-th element of $\mathbf{E}_1$ is $\mathbf{h}_{i}^{\text{H}}\mathbf{x}_{j}^{\left( l-1 \right)}$, and the $(i,j)$-th element of $\mathbf{E}_2$ is $\alpha_i\mathbf{h}_{i}^{\text{H}}\mathbf{x}_{j}^{\left( l-1 \right)}$.
\end{proposition}

\begin{IEEEproof}
To briefly summarize, by considering $\mathbf{X}^{\left(l\right)}$ as the learned precoder $\mathbf{V}$ in the $l$-th layer, we substitute $\mathbf{V}$ with $\mathbf{X}^{\left(l\right)}$ in the objective function \eqref{E:obj}. We then substitute \eqref{adaptive_matrix_form1} into \eqref{E:obj}. Next, we perform a Taylor expansion to \eqref{E:obj} with respect to the signal-to-interference and noise ratio. From the result, we can observe that
     $\overline{\mathcal{G}}^{\left( l-1 \right)}(\cdot)$ depends solely on the terms  $\mathbf{h}_{i}^{\text{H}}\mathbf{x}_{j}^{\left( l-1 \right)}$ and ${{\alpha }_{i}}\mathbf{h}_{i}^{\text{H}}\mathbf{x}_{j}^{\left( l-1 \right)}$. The detailed proof is omitted here due to lack of space.
\end{IEEEproof}

Upon substituting \eqref{E:Gbar} into \eqref{adaptive_matrix_form1}, we obtain the nonlinear weight function as
\begin{align} \label{adaptive_matrix_form}
  {{\mathbf{X}}^{\left( l \right)}}={{\mathbf{X}}^{\left( l-1 \right)}} {{ \mathsf{\mathcal{G}}^{\left( l \right)}\left( \mathbf{E}_1, \mathbf{E}_2 \right)}}.
\end{align}
We can verify that the weight function in \eqref{adaptive_matrix_form} satisfies the unitary equivariant property. Specifically, by transforming ${{\mathbf{X}}^{\left( l-1 \right)}}$ and $\mathbf{H}$ into ${\mathbf{U}{\mathbf{X}}^{\left( l-1 \right)}}$ and $\mathbf{U}\mathbf{H}$, we find that $\mathbf{E}_1$ and $\mathbf{E}_2$ remain unchanged. Consequently, the weight matrix $\mathsf{\mathcal{G}}^{\left( l \right)}\left(\mathbf{E}_1, \mathbf{E}_2 \right)$ also remains unchanged. Thus, by substituting ${{\mathbf{X}}^{\left( l-1 \right)}}$ with ${\mathbf{U}{\mathbf{X}}^{\left( l-1 \right)}}$, we can conclude that the output ${{\mathbf{X}}^{\left( l \right)}}$ is transformed into ${\mathbf{U}{\mathbf{X}}^{\left( l \right)}}$ accordingly.

Next, our task is to ensure that the weight function in \eqref{adaptive_matrix_form} satisfies the PE property with respect to the indices of UEs. By permutating the order of UEs' indices with a permutation matrix $\mathbf{\Pi}^T$, the PE property requires
\begin{equation}\label{PEPE}
    {{\mathbf{X}}^{\left( l \right)}}\mathbf{\Pi}^T={{\mathbf{X}}^{\left( l-1 \right)}}\mathbf{\Pi}^T{{ \mathsf{\mathcal{G}}^{\left( l \right)}\left( \mathbf{\Pi}\mathbf{E}_1\mathbf{\Pi}^T, \mathbf{\Pi}\mathbf{E}_2 \mathbf{\Pi}^T\right)}}.
\end{equation}
Substituting \eqref{adaptive_matrix_form} into \eqref{PEPE} and considering that the obtained condition should hold for arbitrary ${{\mathbf{X}}^{\left( l \right)}}$, we can derive the condition for the function $\mathsf{\mathcal{G}}^{\left( l \right)}(\cdot)$ as
    \begin{equation}\label{jointPE}
        {{\mathbf{\Pi }}}{{ \mathsf{\mathcal{G}}^{\left( l \right)}\left( \mathbf{E}_1, \mathbf{E}_2 \right)}}\mathbf{\Pi }^T={{ \mathsf{\mathcal{G}}^{\left( l \right)}\left( \mathbf{\Pi}\mathbf{E}_1\mathbf{\Pi}^T, \mathbf{\Pi}\mathbf{E}_2 \mathbf{\Pi}^T\right)}}.
    \end{equation}

\subsubsection{\underline{Sub-DNN for learning $\mathsf{\mathcal{G}}^{\left( l \right)}(\cdot)$}}
We employ a dedicated sub-DNN to parameterize the function $\mathsf{\mathcal{G}}^{\left( l \right)}(\cdot)$, ensuring that the sub-DNN satisfies the PE property as given by \eqref{jointPE}.

The matrices $\mathbf{E}_1$ and $\mathbf{E}_2$ can be viewed as features of two feature channels for the input layer. For notational simplicity, we take one feature channel as an example. The output of each hidden layer can also consist of multiple feature channels. We also consider only one of them here. Extending the sub-DNN to accommodate multiple feature channels is straightforward, similar to traditional DNNs.

Let the output of each layer have the same dimensions as $\mathbf{E}_1$ or $\mathbf{E}_2$, denoted as $\mathbf{Y}^{(q)}\in\mathbb{C}^{K\times K}$ for the $q$-th layer. Considering a linear weight function for the sub-DNN and temporally ignoring the activation functions, the input-output relationship for the $q$-th layer can be expressed~as
\begin{equation}
   \overrightarrow{\mathbf{Y}}^{(q)} = \mathbf{W}_{\mathcal{G}}^{(q)}\overrightarrow{\mathbf{Y}}^{(q-1)},
\end{equation}
where $\mathbf{W}_{\mathcal{G}}^{(q)}\in\mathbb{C}^{K^2\times K^2}$ is the weight matrix of the $q$-th layer.
The PE property given by \eqref{jointPE} requires that $\mathbf{W}_{\mathcal{G}}^{(q)}$ satisfies the following condition for all permutation matrix $\mathbf{\Pi}$
\begin{align} \label{E:condition33}
  (\mathbf{\Pi}\otimes\mathbf{\Pi})\mathbf{W}_{\mathcal{G}}^{(q)} = \mathbf{W}_{\mathcal{G}}^{(q)}(\mathbf{\Pi}\otimes\mathbf{\Pi}).
\end{align}
The structure of $\mathbf{W}_{\mathcal{G}}^{(q)}$ that satisfies \eqref{E:condition33} has been investigated in \cite{jointPE}, which contains 15 free parameters. To reduce the number of free parameters, we consider a strengthened condition as follows
\begin{align} \label{E:condition34}
  (\mathbf{\Pi}_1\otimes\mathbf{\Pi}_2)\mathbf{W}_{\mathcal{G}}^{(q)} = \mathbf{W}_{\mathcal{G}}^{(q)}(\mathbf{\Pi}_1\otimes\mathbf{\Pi}_2),
\end{align}
where $\mathbf{\Pi}_1$ and $\mathbf{\Pi}_2$ are two arbitrary permutation matrices. The condition in \eqref{E:condition34} is indeed stronger than \eqref{E:condition33}, as the latter is a special case of the former obtained by setting $\mathbf{\Pi}_1=\mathbf{\Pi}_2$.

The design of a DNN that satisfies the condition in \eqref{E:condition34} was studied in~\cite{edgegnn}, where an Edge-GNN was proposed with only 3 free parameters in each layer. Therefore, we employ the Edge-GNN to learn the function $\mathsf{\mathcal{G}}^{\left( l \right)}(\cdot)$.

\subsection{Design of Function $\sigma^{(l)}(\cdot)$}
In the previous subsections, we designed the weight function $\psi^{\left( l \right)} \left( \cdot  \right)$ that satisfies the joint unitary and permutation equivariant property. To make the whole DNN satisfy the joint property, it is still necessary to design the activation function $\sigma^{(l)}(\cdot)$. We define $\sigma^{(l)}(\cdot)$ as
\begin{align}
  {{\sigma }}\left( {\mathbf{X}}^{\left( l \right)} \right)=\frac{\mathbf{\mathbf{X}}^{\left( l \right)}}{1+{{\left\| \mathbf{\mathbf{X}}^{\left( l \right)} \right\|}^{2}}}.
\end{align}
We can verify that this choice of $\sigma^{(l)}(\cdot)$ satisfies the joint equivariant property.

For the sub-GNN, which only satisfies the PE property, traditional point-wise activation functions can be applied.

\section{Simulation Results}

\begin{figure*}
    \centering
    \begin{minipage}[t]{0.32\linewidth}	
        \subfigure[$\text{SNR}=0\text{~dB}$]{
        \includegraphics[width=0.96\linewidth]{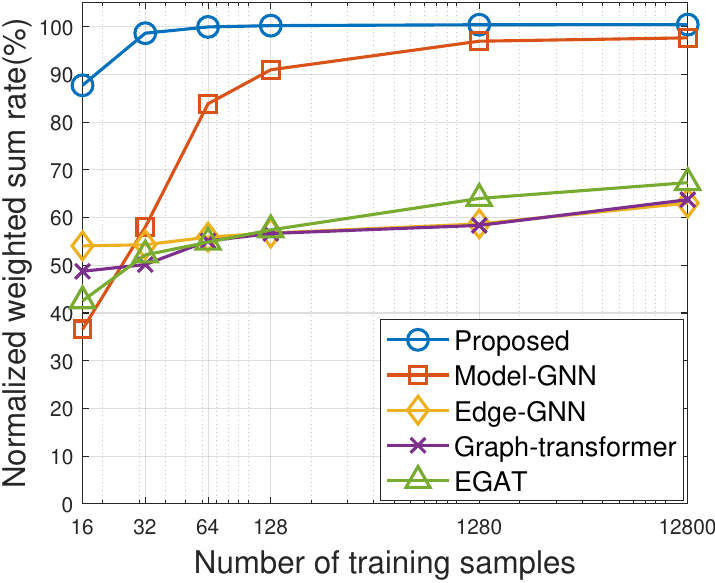}}
    \end{minipage}
    \begin{minipage}[t]{0.32\linewidth}	
        \subfigure[$\text{SNR}=5\text{~dB}$]{
        \includegraphics[width=0.96\linewidth]{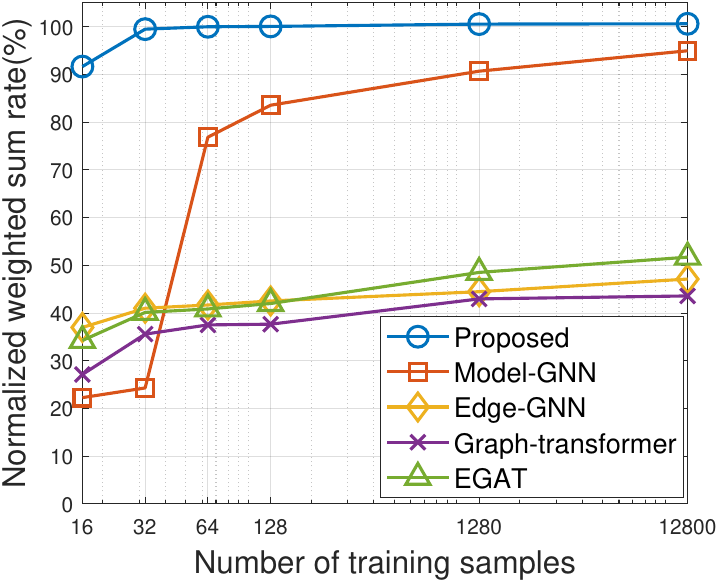}}
    \end{minipage}
    \begin{minipage}[t]{0.32\linewidth}	
        \subfigure[$\text{SNR}=10\text{~dB}$]{
        \includegraphics[width=0.96\linewidth]{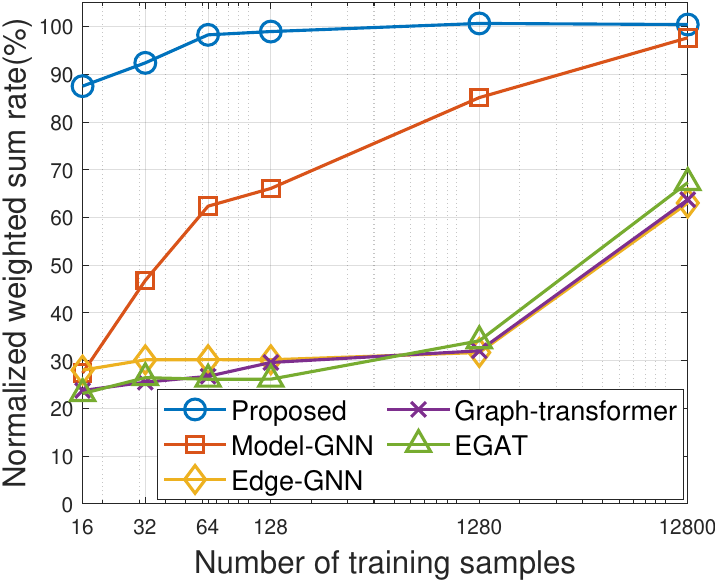}}
    \end{minipage}
    \vspace{0mm}
    \caption{Learning performance at different cell-edge SNRs with $K=16$ and $N=32$.}
    \label{fig:learning_performance}
    \vspace{-4mm}
\end{figure*}

In this section, we evaluate the performance and complexity of the proposed DNN, and compare it with the following baseline numerical algorithm and existing DNNs.
\begin{itemize}
    \item \textbf{WMMSE}: This is a near-optimal numerical algorithm for solving the WSRM problem~[2].
    \item \textbf{Model-GNN}: This is a model-driven GNN proposed in \cite{model_gnn}, leveraging the Taylor's expansion of matrix pseudo-inverse.
    \item \textbf{Edge-GNN}: This GNN was proposed in \cite{edgegnn} satisfies the PE property regarding the indices of both UEs and antennas.
    \item \textbf{Graph-transformer}: This graph transformer architecture was proposed in \cite{Graph_transformer_for_wireless_policy} for precoder learning, involving the attention mechanism as designed in~\cite{Transformer}.
    \item \textbf{EGAT}: This GNN, proposed in \cite{EGAT}, shows improved performance than the traditional GAT introduced in \cite{GAT}.
\end{itemize}

In the simulations, UEs are randomly distributed within a marco cell with a radius of $200$~m radius. The small-scale channels follow the Rayleigh fading model. The large-scale channels follow the 3GPP specifications in \cite{3gppTR38901}, where the path loss at 1 meter is set to 13.54 ~dB and the path loss exponent is set to 39.08. Noise power is adjusted based on the considered cell-edge signal-to-noise ratio (SNR).

To evaluate  fairness among UEs, we generate the channels for each UEs over 20 time slots, where the correlation coefficient between channels in adjacent time slots is set to 0.9. After the simulation of each time slot, we compute the average sum rate over past time slots for each UE and set its inverse as the weight of that UE.
The weighted sum rates achieved by all DNNs are normalized by those achieved by the WMMSE algorithm.
Additionally, we present the cumulative distribution function (CDF) curves of the average rate over 20 time slots to illustrate the performance and fairness among UEs achieved by different methods.



For all DNNs, we set the batch size to 256 and the learning rate to 0.005. The number of training samples is 12,800, unless otherwise specified.  All results are obtained on a personal computer with an Intel CoreTM i9-13900KF CPU and an Nvidia RTX 4080 GPU.

\subsection{Performance Comparison}

\begin{figure}
	\centering
	\includegraphics[width=0.84\linewidth]{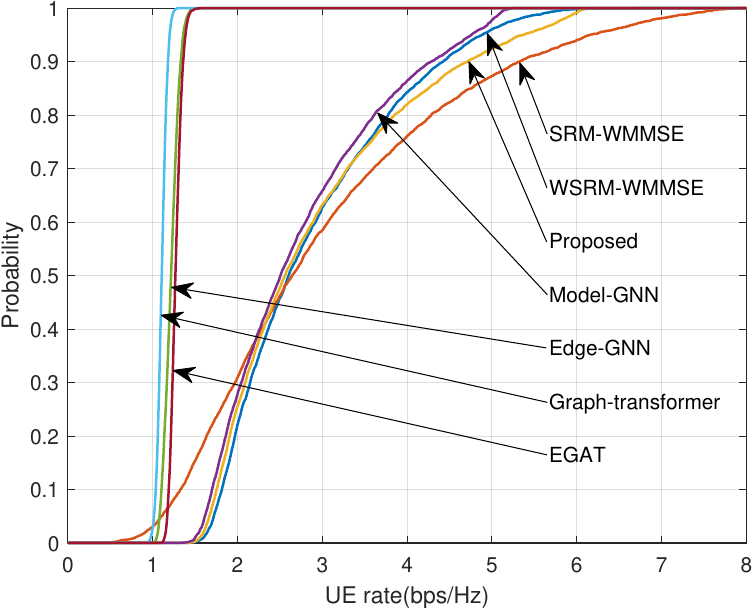}
	\vspace{-1mm}
	\caption{CDF of the average rates of UEs.}
	\label{fig:CDF_learning_performance}
	\vspace{-2mm}
\end{figure}

\begin{figure}
	\centering
	\includegraphics[width=0.84\linewidth]{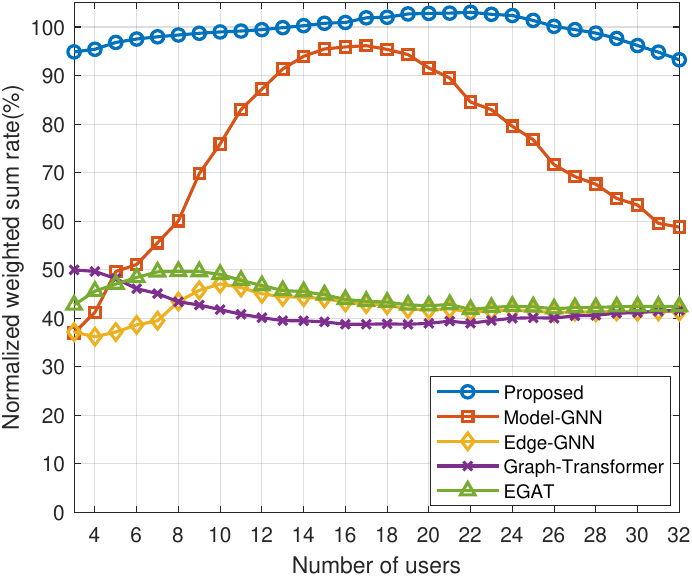}
	\vspace{-1mm}
	\caption{Generalization performance to the number of UEs with $N=32$.}
	\label{fig:generelization_ue}
	\vspace{-3mm}
\end{figure}

In Fig. \ref{fig:learning_performance}, we compare the learning performance of DNNs under different number of training samples and cell-edge SNRs. It can be observed that as the SNR increases and the number of training samples decreases, the learning performance of the proposed DNN does not decline significantly. Even with only 16 training samples and an SNR of 10~dB, the proposed DNN achieves  $87\%$ of the performance of WMMSE. Model-GNN achieves comparable performance when the training sample size is sufficient and the SNR is low, but the gap from the proposed DNN increases as the training sample size decreases and the SNR increases. The other three DNNs were unable to achieve good learning performance in all testings, and their performance further deteriorates as the number of training samples decreases and the SNR increases.

In Fig. \ref{fig:CDF_learning_performance}, we compare the learning performance through the CDF curves of the average rates of UEs, where $N=32$, $K=16$, and $\text{SNR}=5\text{~dB}$. The WMMSE algorithm is applied to solve both the SRM and WSRM problems. As expected, the latter yields better fairness among UEs. The CDF curves of the proposed DNN and Model-GNN are very close to the WMMSE algorithm for the WSRM problem, whereas the other three DNNs exhibit poor performance and fairness.

In Fig. \ref{fig:generelization_ue}, we compare the generalization performance of DNNs to the number of UEs. The DNNs are trained with $N=32$ and $K=16$ at $\text{SNR}=5\text{~dB}$, and tested with $K$ ranging from 3 to 32. It can be seen that compared to the other DNNs, the proposed DNN has the best generalization performance and achieves a normalized sum rate of no less than $90\%$ within the testing range. Model-GNN shows significant performance degradation when the number of UEs is either small or large, while the other three DNNs maintain the performance below~$50\%$.

\begin{figure}
    \centering
    \includegraphics[width=0.84\linewidth]{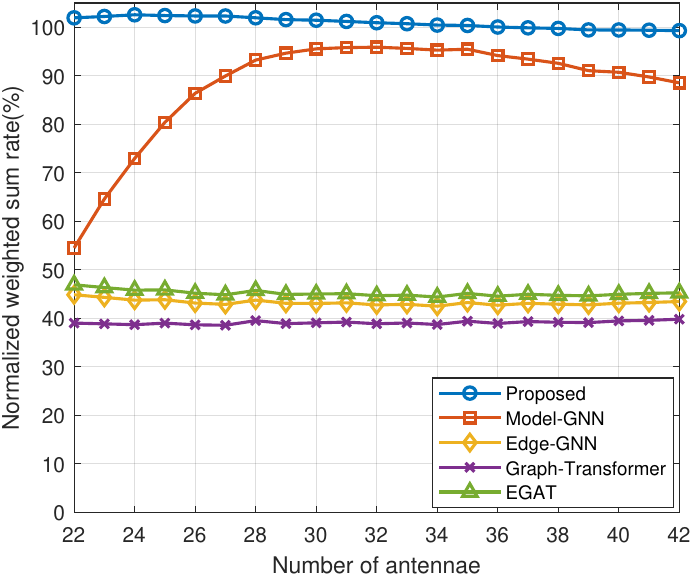}
    \vspace{0mm}
    \caption{Generalization performance to the number of antennas with $K=16$.}
    \label{fig:generelization_tx}
    \vspace{0mm}
\end{figure}

In Fig. \ref{fig:generelization_tx}, we compare the generalization performance of DNNs to the number of antennas. The DNNs are trained with $N=32$ and $K=16$ at $\text{SNR}=5\text{~dB}$, and tested with $N$ ranging from 22 to 42. It can be seen that compared to the other DNNs, the proposed DNN achieves the best generalization performance. Model-GNN exhibits good generalization performance for large $N$, but an obvious degradation for small $N$. The other three DNNs perform poorly across all tested numbers of~antennas.

\begin{figure}
	\centering
	\includegraphics[width=0.84\linewidth]{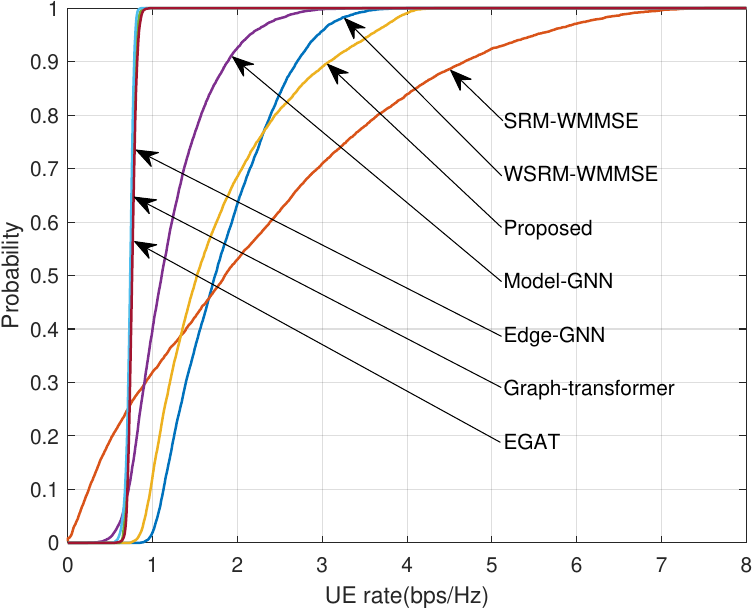}
	\caption{CDF of the average rates of UEs with generalization.}
	\label{fig:CDF_generelization_performance}
	\vspace{-2mm}
\end{figure}

In Fig. \ref{fig:CDF_generelization_performance}, we further evaluate the generalization performance through CDF curves. The DNNs are trained with $N=32$ and $K=16$ at $\text{SNR}=5\text{~dB}$, and tested with $K=30$. It can be seen that when the number of UEs changes from 16 to 30, the proposed DNN still performs closely to the WMMSE algorithm for the WSRM problem. The gap between Model-GNN and the proposed DNN increases compared to the results without generalization as shown in Fig.~\ref{fig:generelization_ue}.

\begin{table*}
    \centering
    \caption{Training Complexity And Inference Time.}\label{table: complexity}
    \footnotesize
    \begin{threeparttable}
    \begin{tabular}{c|c|c|c|c|c}
            \hline\hline
               & Proposed & Model-GNN & Edge-GNN & Graph-transformer & EGAT \\
            \hline
            Sample complexity & 25 & 2000 & $>$12800${}^{*}$ & $>$12800${}^{*}$ & $>$12800${}^{*}$ \\
            \hline
            Time complexity (GPU) & 3.293s & 67.617s & $>$675.848s & $>$1605.791s & $>$6688.557s \\
            \hline
            Space complexity  & 30853 & 51840 & $>$593674 & $>$450304 & $>$152840 \\
            \hline
            Inference time (GPU) & 2.155ms & 2.145ms & $>$1.352ms & $>$4.468ms & $>$3.186ms \\
            \hline
            Inference time (CPU) & 3.568ms & 3.856ms & $>$4.561ms & $>$18.789ms & $>$44.266ms \\
            \hline\hline
        \end{tabular}
        \begin{tablenotes}
        \item ${}^{*}$: For the case marked by ${}^{*}$, the DNN can not achieve the expected performance of $92\%$ with $12800$ training samples. For these DNNs, we record the time complexity to complete 100 training epochs.
        \end{tablenotes}
    \end{threeparttable}
\end{table*}

\subsection{Complexity Comparison}
In this subsection, we compare the sample complexity, time complexity and space complexity of DNNs, which are defined respectively as the minimal number of training samples, the time used for training, and the number of free parameters required to achieve the expected performance (set to 92$\%$ when trained with $N=32$ and $K=16$ at $\text{SNR}=5\text{~dB}$). The results are shown in Table \ref{table: complexity}.

It is shown that only the proposed DNN and Model-GNN can achieve the expected performance. By leveraging the joint unitary and permutation equivariant property, the proposed method achieves lower sample complexity, time complexity, and space complexity than Model-GNN while having comparable inference time. The time complexity and CPU inference time of Graph-transformer and EGAT are significantly higher than others. This is primarily due to the tensor concatenation operations associated with the multi-head attention mechanism, which have high complexity for large batch sizes and cannot be effectively accelerated by parallel computing on the~CPU.

\section{Conclusions}
In this paper, we studied the learning of precoder for weighted sum rate maximization in multiuser multi-antenna systems. We designed a DNN by leveraging the joint unitary equivariant and permutation equivariant property. We demonstrated that a unitary equivariant DNN using a linear weight function in each layer cannot learn the optimal precoding policy. We then proposed a DNN with a nonlinear weight function and employed a GNN to learn this function by exploiting the permutation equivariant property. Simulation results demonstrated that the proposed DNN outperforms baseline DNNs in both learning and generalization performance with low training and inference complexity.

\bibliography{hm_proportional_fairness}

\end{document}